# Evolution of texture and microstructure during accumulative roll bonding of aluminum AA5086 alloy

*Shibayan Roy, Satyaveer Singh D., Satyam Suwas, S. Kumar, K. Chattopadhyay*

*Abstract: In the present investigation, a strongly bonded strip of an aluminium–magnesium based alloy AA5086 is successfully produced through accumulative roll bonding (ARB). A maximum of up to eight passes has been used for the purpose. Microstructural characterization using electron backscatter diffraction (EBSD) technique indicates the formation of submicron sized (~200–300 nm) subgrains inside the layered microstructure. The material is strongly textured where individual layers possess typical FCC rolling texture components. More than three times enhancement in 0.2% proof stress (PS) has been obtained after 8 passes due to grain refinement and strain hardening*

## 1. Introduction

Aluminium and its alloys are quite extensively used in aerospace and automotive industries where superior strength and sufficient ductility are the prime selection criteria. Amongst various strengthening mechanisms reported, refinement in grain size is the most suitable for strength improvement without compromising in ductility [1]. This has led to the quest of fabricating materials with small grain sizes, in ultrafine grained (UFG, ~0.1–1 m) or even smaller (<100 nm). Severe plastic deformation (SPD) based approaches have been found potential to produce such fine grain sized materials in bulk scale. These processes usually involve application of very large strain, which leads to microstructural refinement through subgrain formation and subsequent continuous dynamic recrystallization [3]. An added advantage is minimal change in initial sample dimensions during successive processing steps in a multistep processing schedule. Different variations of SPD processes, e.g. equal channel angular pressing (ECAP), high pressure torsion (HPT), multi-axial forging (MAF), etc., have been successfully applied to fabricate UFG materials [4–15]. Accumulative roll-bonding (ARB) is one such SPD process which has great potential for industrial scaling up since it does not need forming facilities with large load capacity [16]. ARB process typically consists of bonding two stacked sheets by rolling, sectioning the bonded strip in two halves, again stacking them together and rolling. By repeating this process for several cycles, very high strains can be introduced in the materials. This will in turn lead to significant structural refinement [17]. Sheets and plates can be manufactured industrially due to its feasibility as a continuous process.

Although microstructure and texture evolution in various ARB processed aluminium alloys have been documented previously [18–25], micro-mechanisms that lead to the formation of microstructure and texture has not been fully explored yet. The present research program is aimed at addressing this issue. In this study, ARB of an aluminium–magnesium based alloy AA5086 has been carried out up to 8 passes. Systematic effort has been put forward to identify the mechanisms responsible for microstructure and texture evolution in ARB processed material. Finally, the mechanical properties of 8 pass processed material have been evaluated and correlated with corresponding microstructure and texture.

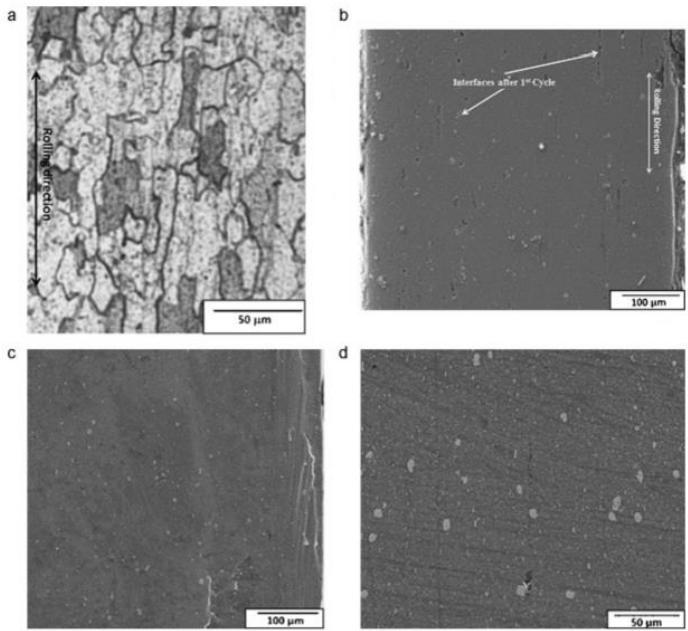

Fig. 1. SEM micrographs showing cross-sectional view of (a) the starting material and (b), (c) and (d) the ARB processed materials after 2, 4 and 6 passes. The rolling direction marked in (b) is also applicable to (c) and (d).

## 2. Experimental procedure

### 2.1. Material and ARB processing

The composition of the AA5086 alloy used for the present study is presented in Table 1. The as-received hot rolled plate (6 mm thickness) was initially cold rolled to a thickness of 0.5 mm and subsequently annealed at 673K (400 ∘C) for 1 h. This cold rolled and annealed material (hereafter referred to as the starting material) was cut into two halves and virgin surfaces were created by cleaning, roughening and degreasing with acetone. These two strips were then stacked with the roughened virgin surfaces facing each other, heated at 573K (300 ∘C) for 5 min, rolled to 50% thickness reduction without any lubrication in a single pass and finally air cooled. Roll-bonded sheet thus produced was further divided into two halves and stacked after roughening the surface. The entire process was repeated up to eight passes to impart a total equivalent strain of 6.4. The number of interfaces thus generated after an ARB pass would be (2n − 1), where n is the number of ARB passes.

### 2.2. Microstructural characterization

Microstructures of the starting and ARB-processed materials after various passes (2, 4, 6 and 8) were characterized by scanning electron microscope (SEM)1 at the transverse plane (RD–ND plane). The observation surfaces were prepared by initial metallo graphic and final electro-polishing 2 using an electrolyte consisting of 78 ml perchloric acid, 90 ml distilled water, 730 ml ethanol and 100 ml butoxyethanol. The starting strips were etched with 0.5% HF solution while the ARB-processed samples were observed without etching in back scattered electron mode. The 8 pass ARB processed material was also characterized using the electron back-scattered diffraction (EBSD) technique3 from the RD–ND plane. An area of 20 m × 10 m was scanned using a step size of 50 nm. [26-30] The orientation data obtained from EBSD scan was analyzed using TSL OIMTM Analysis Software4 considering minimum boundary misorientation to be 2∘. In the EBSD scans, the hit rate was ∼90% and the confidence index (CI) equaled to 0.26 which indicates the accuracy of the analysis. The EBSD orientation data is cleaned using 'grain dilation' method which is an iterative process and only acts on point/pixel that do not belongs to any grains; yet have neighboring points/pixels which are part of certain adjacent grains. Such a point/pixel is either un-indexed during the scan or fails to meet the grain defining criterion.

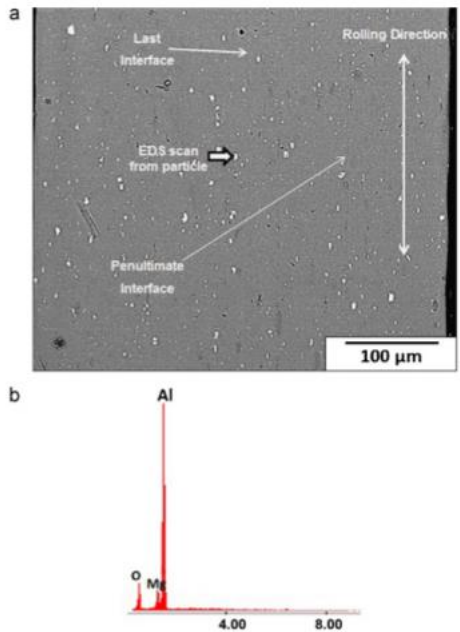

Fig. 2. SEM micrographs showing the cross-sectional view of ARB processed material strip after 8 passes. The rolling direction is marked in the micrograph.

Additionally, a criterion was set up that the CI of the cleaned point/pixel should be <0.1. If majority of the neighboring pixels belong to one single grain, the orientation of the un-indexed pixel will change to match this adjacent grain. In otherwise, the orientation will be selected to match any of the neighboring pixels which belong to the grain with the highest average CI. This process is repeated until each point in the data set becomes the member of a grain.

### 2.3. Texture characterization

Bulk texture of the starting and ARB processed materials were measured using X-ray texture goniometer based on Schulz reflection geometry5 using Cu K radiation at the mid thickness region (RD–TD plane). Four incomplete pole figures, viz., {1 1 1}, {2 0 0}, {2 2 0} and {3 1 1} were measured experimentally. Quantitative texture analysis was carried out by calculating the orientation distribution function (ODF) using LaboTex6 software. Arbitrarily Defined Cells (ADC) algorithm was used and no restriction of specimen symmetry was imposed while calculating the ODFs.

### 2.4. Mechanical testing

Tensile tests at room temperature were carried out on the starting and 8 pass ARB processed materials along three directions (0◦, 45◦ and 90◦ with respect to the RD) in order to assess the mechanical anisotropy in the material owing to texture. From statistical considerations, three specimens were tested for each orientation. Tensile properties were estimated fromtrue stress–true strainplots using the standard procedure. Specimen elongations to failure have been calculated from the difference before and after the test only from the uniform sections of the gauge lengths. Further, specimen elongations are reconfirmed from the uniform part of the true stress–true strain curves.

# 3. Results

### 3.1. Microstructural evolution

Microstructure of the starting material shows a combination of equiaxed (∼50 m diameter) and elongated grains (∼100 m diameter). Representative micrograph is given in Fig. 1a. The ARB processed materials after 2, 4 and 6 passes show sufficient bonding between different layers (Figs. 1b–d). Fig. 2a represents the microstructure of the ARB-processed material after 8 passes with the penultimate interfaces being marked in the micrograph. After 8 passes, the bonding between successive layers are excellent on either side of the last interface up to the extent that the interfaces developed during previous passes are difficult to identify. The interfaces are in fact revealed by the presence of shinning particles of width ∼5–10 m, oriented along the rolling direction. Elemental analysis indicates these particles to be the oxides of aluminium (Fig. 2b).

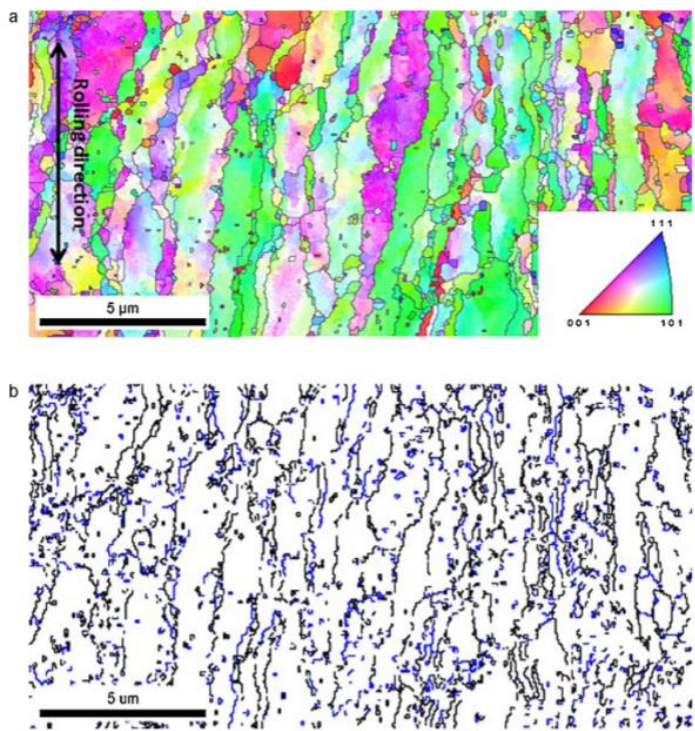

Fig. 3. Inverse pole figure (IPF) map and (b) grain boundary map (low angle grain boundaries with misorientation < 15◦ shown in blue and high angle boundaries with misorientation > 15◦ shown in black) for the 8 passes ARB processed material. The rolling direction shown in (a) is also applicable to (b). (For interpretation of the references to color in this figure legend, the reader is referred to the web version of the article.)

## 3.2. Microstructural investigation by EBSD analysis

A finer microstructural characterization of 8 pass ARB processed materials was carried out by EBSD technique. The inverse pole figure map shows a layered microstructure mostly composed of elongated grains along the rolling direction (Fig. 3a). Lesser but noticeable fraction of equiaxed grains surrounded by high angle boundaries (misorientation > 15◦) are also present in the microstructure, more frequently at the interfaces between successive layers (Fig. 3b). Individual layers are characterized by low angle boundaries and intra-layer orientation gradients, although drastic change in orientation has not been usually observed within a single layer. Adjacent layers are, however, mostly of different orientations. Clearly evident from Fig. 4a that significant proportion (75%) of high angle grain boundaries (HAGB) develops in addition to low angle boundaries (LAGB) after 8 pass of ARB. The maxima in the misorientation distribution occur at ∼5–10◦ for LAGB and at ∼55◦ for HAGB. The grain size distribution in the ARB processed material shows substantial refinement in comparison to the starting material (Fig. 4b). A significant fraction (∼50%) of grains have diameter <100 nm while the average lies within 150–200 nm. [31-35]

### 3.3. Texture evolution

The variation of maximum intensity in (1 1 1) pole figure (m.r.u.) with number of ARB passes suggests that the overall texture in starting material is much weak compared to that for the ARB processed materials (Table 2). Table 3 lists the typical texture components and the corresponding Euler angles that form during deformation (e.g. S, copper and brass) and recrystallization (e.g. cube and Goss) of FCC metals and alloys. In addition, certain shear texture components that are typical for ARB have also been included in the table (e.g. Dillamore). The relevant ODF sections ($\phi2 = 0°$, 45° and 65°, Fig. 5) for the starting and ARB processed materials show the presence of these FCC rolling texture components. In Table 4, the strength of individual texture components (in terms of orientation density maxima f(g), as calculated from the orientation distribution function) for the starting as well as ARB processed materials have been tabulated. The strength of deformation texture components enhances significantly after ARB up to 4 passes. A maximum is observed at this stage and texture weakens thereafter. ARB processed material shows the presence of almost all the deformation components (copper, Dillamore, brass and S). Most noteworthy are copper and Dillamore components. These two components significantly strengthens after 2 passes, reaches the maxima after 4 passes, stays almost constant up to 6 passes and finally decreases after 8 passes of ARB. The strength of S-component follows the same trend up to 4 passes. The Brass component, however, shows continuous increase in strength and becomes stronger after 8 passes, than copper, S and Dillamore components. The strength of recrystallization texture components, e.g. Goss or cube does not significantly vary from starting material during the course of ARB. The Goss component remains almost constant throughout while the cube component initially decreases to a very low value for 2 passes and strengthen again up to 8 passes.

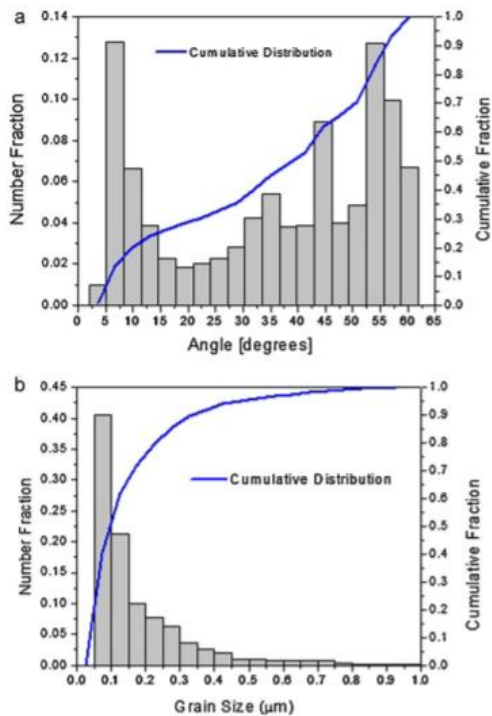

Fig. 4. (a) Misorientation and (b) grain size distribution in the 8 passes ARB processed material.

### 3.4. Mechanical characterization

True stress–true strain plots for the starting material do not exhibit much effect of specimen orientation (Fig. 6a). These curves are characterized with a yield point followed by serrated flow. In contrast, ARB processed material exhibits a large anisotropy in mechanical behavior with respect to specimen orientation (Fig. 6b). The stress–strain curves, however, do not show sharp yield points and neither have they exhibited serrated flow. Instead,these curves display instabilities in the post necking region. More than three times improvement is marked in strength after ARB compared to the starting material (Fig. 7a). However, the strength improvement is accompanied with a sharp decrease in ductility, as evident from Fig. 7b. Much stronger anisotropy in strength has been noted for ARB processed material with the specimen tested along 90◦ possesses the highest yield strength. [36-41] Strength of the specimens tested along 0◦ or 45◦ is almost the same. The ductility, on the other hand, varies with testing direction for both starting and ARB processed material. The amount of elongation increases in the order 0◦ < 45◦ < 90◦ for starting material. After ARB processing, highest elongation is recorded when tensile test is carried out along 45◦ and ductility varies in the order 45◦ > 90◦ > 0◦. The representative micrographs from the fracture surfaces perpendicular to the tensile axis are shown in Fig. 6 inset. For starting material, typical ductile fracture surface is observed which includes numerous equiaxed/hemispheroidal dimples with a shiny fibrous outline and shear zones atthe outside surfaces oftensile specimens (Fig. 6a inset). In case of 8 pass ARB processed material, the fracture surfaces primarily exhibit debonding of layers, delamination being most prominent along the final interface (Fig. 6b inset). Similar to the starting material, the fracture surface of the ARB processed material also possesses equiaxed dimple between the interfaces.

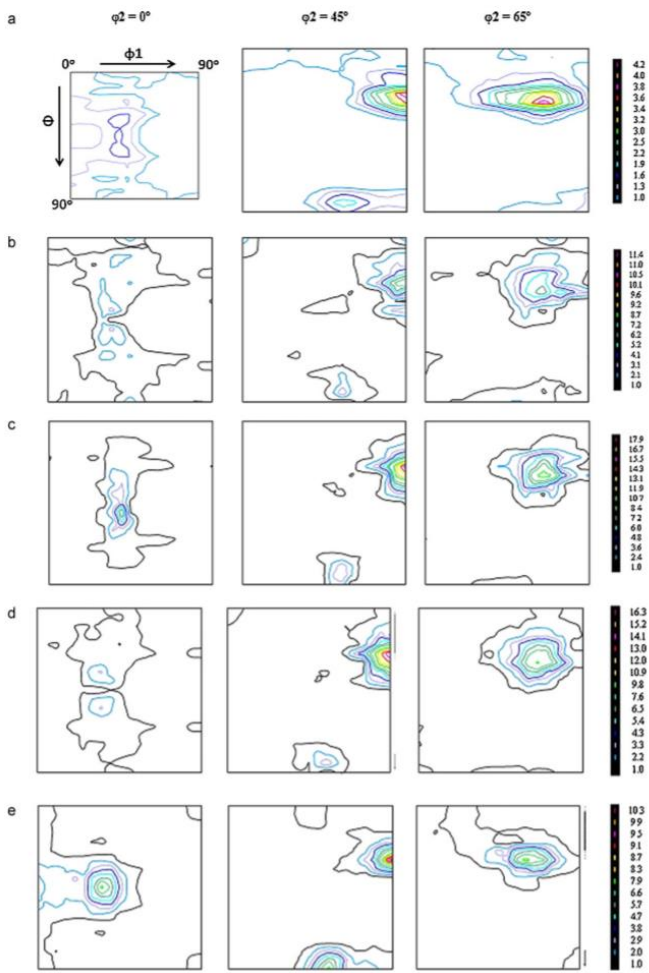

Fig. 5. Selected ODF sections ($\phi2 = 0$◦, 45◦ and 65◦) of (a) starting material and after (b) 2 passes, (c) 4 passes, (d) 6 passes, and (e) 8 passes of ARB.

# 4. Discussion

4.1. Microstructural evolution during ARB

4.1.1. General microstructural features Abundant presence of LAGB inside elongated grains after 8 passes of ARB is normal for any SPD process and indicates presence of subgrains in the microstructure (Fig. 4a). The main deformation mechanism during SPD process is crystallographic slip associated with organization of dislocations into random and geometrically necessary boundaries [3]. As the strain increases, random boundaries form substructures while geometrically necessary boundaries give rise to deformation bands. DuringARB, deformation bands lead to the formation of shear bands under severe straining. This process eventually gives rise to a layered microstructure separated by high angle boundaries [12]. The equiaxed grains that are bounded by high angle boundaries at the inter-layer interfaces possibly result from continuous dynamic recrystallization. Shear bands that form in the microstructure during successive ARB operations can act as a potential nucleation sites for such recrystallization.

4.1.2. Mechanism of microstructural refinement Finer examination of microstructural features via EBSD analysis reveals that the grain size after 8 passes ARB is ~200 nm which is almost two orders of magnitude lower than the starting material (~50–60 m)(Fig. 4b). Grainrefinementduring SPD occurs through the generation and arrangement of dislocations in regular substructures separated by high angle boundaries (HAGB). The process can be visualized as a continuous rearrangement of free dislocations from planar arrays, to microbands, to elongated cells structures and finally to subgrains with increasing plastic strain [18,19]. A detailed description of microstructural evolution and sub-structure formation during severe plastic deformation processes like ARB, can be found in Refs. [42-48]. Since the equiaxed subgrains need to accommodate large rotation of neighboring regions, they are usually surrounded by high angle boundaries. Subgrain rotation might result from either the activation of multiple slip systems in different parts of the grains or through strain subdivision process in which different regions of the same grain undergo varying strains and rotate relative to each other.

4.1.3. Substructure formation Development of sub-structural features inside individual layers can be understood by plotting misorientation along a line on point-to-origin as well as point-to-point basis. For example, misorientation variation for two adjacentlayers can be considered (Fig. 8). The point-to-origin misorientation compares the orientation of each point with respect to the starting point while the point-topoint misorientation compares the misorientation of each point with respect to the previous point. In case of the first layer (line 1), total misorientation on point-to-origin basis is ~5∘. Average point-to-point misorientation is comparatively low (~1∘) although sudden increase up to ~4∘ can be noticed. It is to be noted that the misorientation within 1–2∘ is below the limit of resolution for EBSD technique (2∘). The difference between two misorientation bases suggests that misorientation builds up within this layer via a subtle additive process during deformation. Grain elongation during ARB needs to accommodate severe shear deformation without instability which is possible via micro-band formation [18]. Microbands eventually lead to a continuous increase in misorientation on pointto-origin basis across an individual layer. For the second layer (line 2), both point-to-origin and pointto-point misorientation shows large variation across the layer. Occasionally, point-to-point misorientation reaches ~3–4∘ and as high as 8∘ in some locations. Sudden increase in intra-layer misorientation corresponds to subgrain boundaries. Low misorientation regions, on the other hand, are related to cell interiors with relatively fewer dislocations. It is, therefore, quite evident in the first layer (line 1)thatthe final microstructure is characterized by deformation microband wherein for the second layer (line 2), subgrains are formed. This difference can be explained on the basis of respective starting orientations. During the tensile deformation of polycrystalline aluminium, the grains with <001> tensile axis orientation showed equiaxed cell structure with crystallographic dislocation boundaries parallel to the primary slip planes [49-51]. On the other hand, layered cell structure with non-crystallographic (dislocation walls at a large angle with the primary slip plane) dislocation boundaries forms in grains with <111> tensile axis. Although, multiple slip activates for both the orientations, a large number of slip systems (at least eight) possess similarly high Schmidt factors for <001> tensile axis orientation. On the contrary, in grains with <111> tensile axis orientation, hardly two slip systems possess high and similar Schmidt factor. The sub-structure formation in a polycrystalline material is thus highly orientation dependent. The situation is further complicated since multiple slip systems are needed in order to meet five independent slip system criterion [52-58]. It is possible that the deformation of certain regions of the grain is limited to only the primary slip, although the von Mises' criterion is satisfied globally through poly-slip at other regions, especially near the grain boundaries. Morii et al. [21] have shown that shear band formation is strongly supported by the presence of subgrains which effectively hinders homogeneous glide of sessile dislocations. This indicates that shear bands are more likely in the <111> tensile axis orientated layer rather than the <001> tensile axis orientated layer.

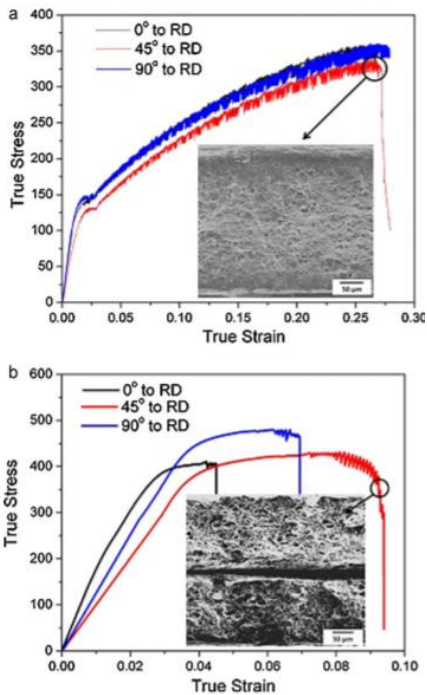

Fig. 6. True stress–true strain curves for (a) the starting material and (b) the ARB processed material after 8 pass at various angles (0◦, 45◦ and 90◦)to rolling direction. (a and b) Insets showing representative fractographs from the starting and ARB processed materials, both tested along 45◦ to the rolling direction.

## 4.2. Texture evolution

4.2.1. Overall texture development Results of bulk texture measurement clearly indicate the evolution of characteristic deformation texture during the course of ARB (Fig. 5). Most of the deformation texture components strengthen significantly from the starting material while the inherited components (generally the recrystallization components) are retained during ARB. One of the important conditions for SPD generated microstructure is the avoidance of discontinuous dynamic recrystallization [24]. The strength invariance for recrystallization components suggests that intermediate annealing does not lead to discontinuous recrystallization (static or dynamic) during ARB. Instead, it facilitates recovery processes in the material prior to ARB. This understanding provides an importantinput regarding the appropriateness of the processing conditions. Individual strength of deformation texture components (e.g. copper, S and brass) increases with number of ARB passes, so does the Dillamore component. Total strength of all the deformation components do not alter much after 4 passes signifying that substructure formation saturates with further straining. After 8th pass, the strength of deformation components decreases from previous exceptfor the brass component. Possible reason for the evolution of stronger brass component could be the formation of shear bands in the microstructure, which are known to be associated with strong brass type texture [59-61].

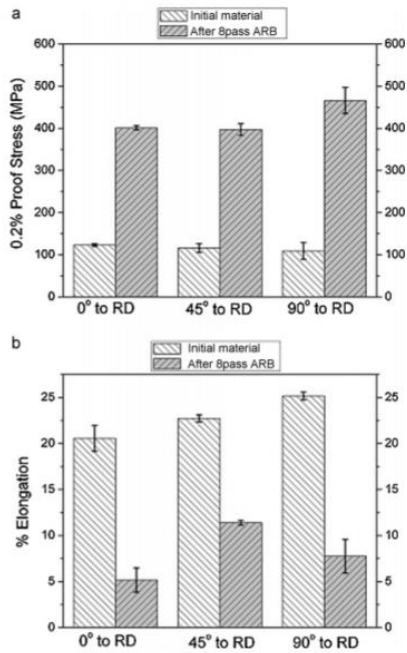

Fig. 7. Mechanical properties of starting material vis-à-vis the 8 pass ARB processed material: (a) 0.2% PS and (b) % elongation to fracture.

### 4.2.2. Micro-texture evolution

In order to correlate texture related issues with microstructure in a finer detail, the respective texture components are mapped in the EBSD generated microstructure (Fig. 9). Primary observation in this regard is that one single component seldom repeats in successive layers. The presence of coarse, unrefined deformation bands in the microstructure during the ARB leads to development of strong texture in individual layers [27]. The S and brass components are quite prevalent in the microstructure. Significant fraction of overlapping copper and Dillamore components appear in the same layer.

4.2.2.1. Origin of copper/Dillamore component. Dillamore component primarily originates from a shear component which occurs in the sheet surface due to frictional shear stress imparted by the rolls. During the next ARB pass, these surfaces appear in the middle of the sheet. Thereafter, rotation of shear component due to interfacial shear during further deformation leads to Dillamore component [12]. This component is closely located to copper component in the orientation space (please refer to Table 3 for exact location of these components), which sometimes makes them difficult to distinguish. Layers characterized by copper/Dillamore component occur at regular interval of ∼2–3 m in the microstructure which is in similar range to the individual layer thickness after 8 passes of ARB. This indicates that Dillamore component is indeed related to surface friction and interface shear during deformation. With further straining, shear band forms in the microstructure transforming the intermediate Dillamore component to copper component as the stable end orientation [28]. Important to note that copper/Dillamore component form exactly in the layer prone to subgrain formation and shear banding during deformation (Figs. 8 and 9).

4.2.2.2. Evolution of brass component. Brass component mostly forms in layers which are adjacent to layers holding copper/Dillamore components. As mentioned in the preceding subsection, formation of copper/Dillamore component is related to shear banding in these layers. Shear banding in a homogenously deforming crystal requires activation of single slip which reduces the amount of latent hardening and hence the total amount of plastic work during deformation [29]. The planar slip promotes brass component so that they mostly occur adjacent to copper/Dillamore layers [25,26].

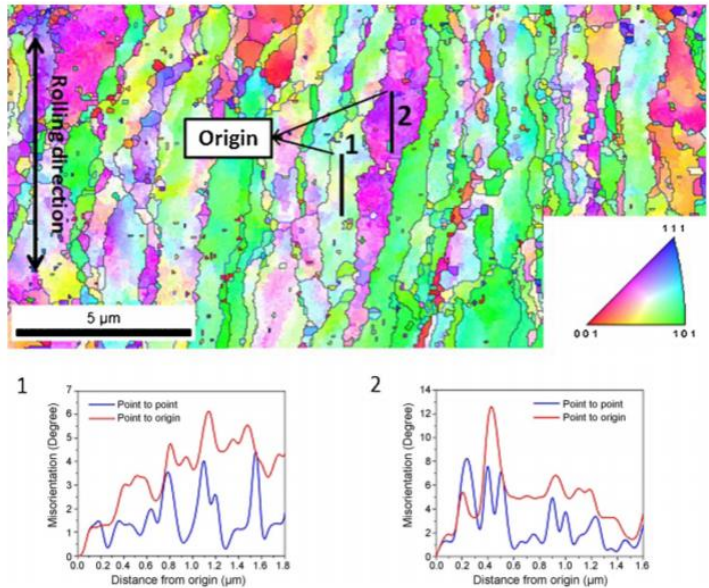

Fig. 8. Change in misorientation inside individual layers in the 8 pass ARB processed material. The rolling direction is marked alongside. The plotted boundaries compose of both LAGB and HAGB.

4.2.2.3. Origin of cube and Goss components. Goss component is present in the microstructure within fine equiaxed grains in the brass oriented grains. Goss component preferably nucleates at shear bands and usually subjugates the brass component [30,31]. Shear bands in the present study form after large straining (8 passes) which facilitate formation of Goss component. Consequently this component strengthens in the bulk texture. Cube oriented grains, on the other hand, nucleate at transition bands or can result from a preferred growth relationship (38° around <111>) with the S-component [30,32]. Subsequently, the cube component forms by the rotation of copper oriented grains towards the    fiber presentinthe deformedmaterial. Themicrostructure indeed shows cube oriented grains betweenthe elongated grains characterized by brass component and S-component. The presence of microbands within these S-oriented grains (their origin has been discussed in Section 4.1.3), provides required nucleation sites for cube oriented grains.

4.3. Mechanical behavior of ARB processed materials

Serrated flow or Portevin-Le Chatelier (PLC) effect in the stress–strain curve for starting materials can be attributed to dynamic strain ageing [33]. The serrations initiate at the onset of plastic deformation and attains a value lower than the uniform stress level (type-C serrations) [34]. Kumar have observed a transition from type-B to type-C serrations at room temperature in case of an Al–5Mg alloy. The onset strain for such serrations was noted to be nearly zero [35]. The magnitude of serrations in the present study increases with strain which is expected according to the dynamic strain ageing mechanism [36]. The flow curves for 8 pass ARB processed material neither exhibit a yield point nor serrated flow. This can be attributed to the fine grain size in the ARB processed material. It has been reported that PLC effect in AA5088 alloy weakens on reducing the grain size from 61 to 35 m and disappears altogether on further reduction to 13 m [37]. Instabilities in the post necking region for ARB processed material indicate continuous local fracture of layers before the final fracture could initiate along the last interface. Strength improvement after ARB can primarily be attributed to grain refinement, although significant contribution is expected from large redundant deformation in this material. Eizadjou et al. [12] have confirmed that strain hardening and sub-grain formation are mainly responsible for strength improvement up to the 4th pass and grain refinement dominates afterwards. The presence of a hard surface layer due to wire brushing may also lead to strength improvement to some extent [38].

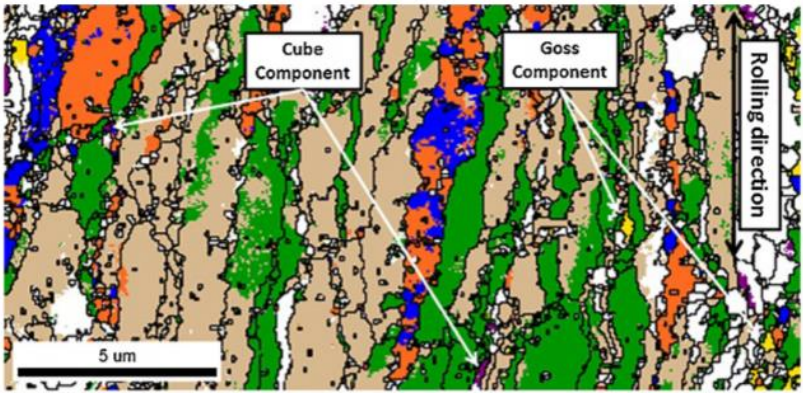

Fig. 9. The EBSD generated map showing the distribution of various texture components in the microstructure of ARB processed material. The color coding and rolling direction are given alongside. The plotted boundaries compose of both LAGB and HAGB.

Decrease in ductility for ARB processed material is primarily due to de-bonding/delamination of layers at an early stage of deformation leading to premature fracture as shown in Fig. 6b inset. It is to be mentioned thatthe interfaces for the specimen tested in allthree directions (0∘, 45∘ and 90∘ to RD) were originally parallel to the loading direction and unlikely to de-bond under tensile loading. It is possible that interfacial shear develops between successive layers once the instability starts at the necked region. Under this condition, the stress state modifies to tri-axial and can lead to debonding of the weakest interface that is located at the mid-thickness of the specimens. This is evident from the fractographs shown in Fig. 6b inset. Such interface debonding has been frequently observed during tensile testing of ARB processed materials with interfaces being originally parallel to the loading direction [12,39]. Strain hardening due to redundant deformation does not contribute much in ductility reduction because of the positive effect from submicron grains in the microstructure [39]. However, presence of dislocation substructure can restrict uniform elongation and result in early initiation of plastic instabilities (e.g. necking). ARB processed material fails largely due to shear stresses prior to complete debonding of individual layers. The starting material, on the other hand, follows normal strain hardening behavior and instabilities occur after a large amount of deformation. The observed anisotropy in strength and ductility is quite expected for ARB processed material due to strong deformation texture formation [40]. Comparatively weak texture in the starting material ensures uniform mechanical response irrespective of test directions. Optimal property combinations for ARB processed material are obtained when tested along 45∘. This is related to the balance between various texture components which behaves as either 'hard' or 'soft' orientations with respectto the applied tensile stress.

## 5. Conclusions

The alloyAA5086 was accumulatively roll-bonded up to 8 passes in the present work. Salient observations from microstructural and crystallographic texture studies and mechanical property evaluation have led to the following conclusions:

(i) A good bonding between successive layers is achieved after ARB in every pass emphasizing the appropriateness of applied processing schedule.

(ii) ARB processed material after 8 passes shows layered microstructure containing elongated and equiaxed grains. The average grain size lies within ∼200–300 nm.

(iii) Significant substructure formation has been identified inside the layered microstructure. Substructures contain subgrains surrounded by LAGB or deformation microbands depending on the starting orientation. Substructure formation finally induces shear banding in the course of deformation.

(iv) Evolution of characteristic deformation texture is observed during ARB. Individual layers in the microstructure form stable texture bands.

(v) The inter-layer shear gives rise to Dillamore component. A strong brass component evolves after 8 passes because of shear banding in the microstructure. Ease of recrystallization at the shear bands maintains the strength of recrystallization components after 8 passes.

(vi) ARB processed material showed strength improvement after 8 passes due to grain refinement and strain hardening. Simultaneous decrease in ductility is primarily due to interface debonding. The mechanical behavior is highly anisotropic owing to the strong texture after ARB.